\documentclass[superscriptaddress,reprint,prb]{revtex4-1}
\usepackage{graphicx}
\usepackage[margin=0.7in]{geometry}
\usepackage{amsmath}
\usepackage{hyperref}
\usepackage{multirow}

\let\hatOrig\hat
\renewcommand{\vec}[1]{\boldsymbol{\mathbf{#1}}}
\renewcommand{\hat}[1]{\boldsymbol{\mathbf{\hatOrig{#1}}}}
\renewcommand{\Im}{\operatorname{Im}}
\renewcommand{\Re}{\operatorname{Re}}
\newcommand{\D}{\mathrm{d}}
\newcommand{\sub}[1]{\ensuremath{_{\textrm{#1}}}} \newcommand{\super}[1]{\ensuremath{^{\textrm{#1}}}} \newcommand{\sci}[2]{\ensuremath{#1 \times 10^{#2}}} 

\newcommand{\narrowfig}[1]{\centering{\includegraphics[width=3.3in]{#1}}}

\newcommand{\JCAP}{Joint Center for Artificial Photosynthesis, California Institute of Technology, Pasadena CA}
\newcommand{\MSC}{Materials and Process Simulation Center, California Institute of Technology, Pasadena CA}
\newcommand{\Watson}{Thomas J. Watson Laboratories of Applied Physics, California Institute of Technology, Pasadena CA}
\newcommand{\NGnext}{NG NEXT, 1 Space Park Drive, Redondo Beach CA}

\begin{document}

\title{\emph{Ab initio} phonon coupling and optical response of hot electrons in plasmonic metals}

\author{Ana M. Brown}\affiliation{\Watson}
\author{Ravishankar Sundararaman}\affiliation{\JCAP}
\author{Prineha Narang}\affiliation{\Watson}\affiliation{\JCAP}\affiliation{\NGnext}
\author{William A. Goddard III}\affiliation{\JCAP}\affiliation{\MSC}
\author{Harry A. Atwater}\affiliation{\Watson}\affiliation{\JCAP}

\date{\today}

\begin{abstract}
Ultrafast laser measurements probe the non-equilibrium dynamics
of excited electrons in metals with increasing temporal resolution.
Electronic structure calculations can provide a detailed microscopic
understanding of hot electron dynamics, but a parameter-free description of
pump-probe measurements has not yet been possible, despite intensive research,
because of the phenomenological treatment of electron-phonon interactions.
We present \emph{ab initio} predictions of the electron-temperature dependent
heat capacities and electron-phonon coupling coefficients of plasmonic metals.
We find substantial differences from free-electron and semi-empirical estimates,
especially in noble metals above transient electron temperatures of 2000~K,
because of the previously-neglected strong dependence of
electron-phonon matrix elements on electron energy.
We also present first-principles calculations of the electron-temperature
dependent dielectric response of hot electrons in plasmonic metals,
including direct interband and phonon-assisted intraband transitions,
facilitating complete theoretical predictions of the time-resolved
optical probe signatures in ultrafast laser experiments.
\end{abstract}

\maketitle

\section{Introduction}

Understanding the energy transfer mechanisms during thermal non-equilibrium between
electrons and the lattice is critical for a wide array of applications.
Non-equilibrium electron properties on time scales of 10-100s of femtoseconds
are most efficiently observed with pulsed laser measurement techniques.\cite{
Anisimov, DelFatti, Elsayed1987, Elsayed-Ali, Giri, Hartland, Kaganov}
Laser irradiation of a metal film or nanostructure with an ultrashort
laser pulse pushes the electron gas out of equilibrium; describing the
evolution of this non-equilibrium distribution has been the subject of
intense research for two decades.
A majority of investigations so far employ various approximate models,
typically based on free-electron models and empirical electron-phonon interactions,
to calculate the energy absorption, electron-electron thermalization and
electron-phonon relaxation. \cite{Grimvall, Gan:15, Link, Luo:2013qf, Norris,
Mueller:2014sf, Groeneveld1990, Groeneveld1995, Rethfeld, Sun, JAP2014Yanbao, Leenheer:2014sw}
However, a complete \emph{ab initio} description of the time evolution and optical response of
this non-equilibrium electron gas from femtosecond to picosecond time scales has remained elusive,
especially because of the empirical treatment of electron-phonon interactions.\cite{Lin}

The initial electron thermalization via electron-electron scattering is qualitatively
described within the Landau theory of Fermi liquids.\cite{Landau, Ziman, Ashcroft, Pines}
The subsequent relaxation of the high temperature electron gas with the lattice
is widely described by the two-temperature model (TTM),\cite{Anisimov,Giri,Lin,Sun,Hartland,Kaganov}
given by coupled differential equations for the electron and lattice temperatures, $T_e$ and $T_l$,
\begin{flalign}
C_e(T_e) \frac{d T_e}{d t} &= \nabla \cdot (\kappa_e \nabla T_e) - G(T_e) \times (T_e - T_l) + S(t) \nonumber\\
C_l(T_l) \frac{d T_l}{d t} &= \nabla \cdot (\kappa_p \nabla T_l) + G(T_e) \times (T_e - T_l).
\label{eqn:TTM}
\end{flalign}
Here, $\kappa_e$ and $\kappa_p$ are the thermal conductivities of the electrons and phonons,
$G(T_e)$ is the electron-phonon coupling factor,
$C_e(T_e)$ and $C_l(T_l)$ are the electronic and lattice heat capacities,
and $S(t)$ is the source term which describes energy deposition by a laser pulse.
In nanostructures, the temperatures become homogeneous in space rapidly and
the contributions of the thermal conductivities drop out.
A vast majority of studies, both theoretical and experimental, treat the remaining material
parameters, $G(T_e)$, $C_e(T_e)$ and $C_l(T_l)$,
as phenomenological temperature-independent constants.\cite{PhysRevB.57.11334,
knoesel1998ultrafast, Gavnholt:2009bc, carpene2006ultrafast,
brorson1987femtosecond, Frischkorn:2006py, Harutyunyan:2015nx, Hostetler}

\begin{figure}
\narrowfig{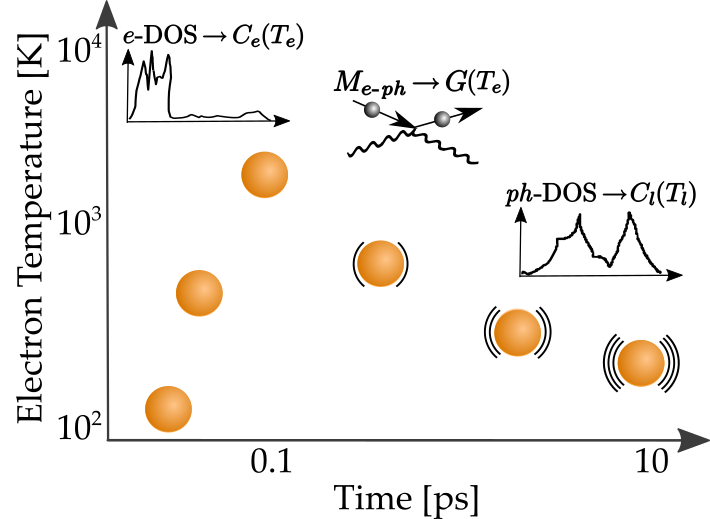}
\caption{Schematic electron and lattice temperature evolution with time
following laser pulse illumination of a plasmonic metal like gold,
along with the relevant material properties that determine this evolution.
The vertical position of the gold atoms on the plot corresponds to electron temperature,
and the vibration marks around the atoms schematically indicate lattice temperature.
We show that both the electron heat-capacity $C_e(T_e)$ (from electronic
density of states (DOS)) that sets the peak electron temperature $T_e$,
and the electron-phonon coupling strength $G(T_e)$ (from electron-phonon
matrix element $M\sub{e-ph}$) that affects the relaxation time of $T_e$,
vary with $T_e$ in a manner sensitive to details of $d$ electrons in noble metals.
Only the lattice heat capacity $C_l(T_l)$, that determines the lattice temperature rise,
does not vary substantially between the detailed phonon DOS and simpler models.
\label{fig:Schematic}}
\end{figure}

Figure ~\ref{fig:Schematic} schematically shows the time evolution of 
the electron and lattice temperatures in a plasmonic metal like gold,
and the role of the \emph{temperature-dependent} material properties.
The electronic density-of-states and the resultant electronic heat capacity $C_e(T_e)$
determine the peak electron temperature $T_e$ reached after electron-electron thermalization.
The electron-phonon matrix elements and the resulting coupling strength $G(T_e)$
determine the rate of energy transfer from the electrons to the lattice,
which along with $C_e(T_e)$ determines the rate of relaxation of $T_e$.
Finally, the phonon density of states and the resulting lattice heat capacity $C_l(T_l)$
determine the rise in lattice temperature $T_l$.

A key challenge in the quantitative application of TTM models is the
determination of these \emph{temperature-dependent} material parameters.
With pulsed lasers, it is possible to absorb sufficient
energy in plasmonic nanostructures to melt the metal
once the electrons and lattice have equilibrated.\cite{Link:1999fk}
The highest electron temperature, $T_e\super{max}$
accessible in repeatable measurements is therefore
limited only by the equilibrated lattice temperature
being less than the melting temperature $T_m$ of the metal,\cite{CRChandbook}
which yields the condition
$\int_{T_m}^{T_e\super{max}} \D T_e C_e(T_e) = \int_{T_0}^{T_m} \D T_l C_l(T_l)$.
Starting at room temperature $T_0 = 300$~K and using our
calculations of the electron and lattice
heat capacities, $C_e(T_e)$ and $C_l(T_l)$,
we find $T_e\super{max} \approx $ 5700, 8300, 7500 and 6700~K
respectively for aluminum, silver, gold and copper. 
For gold and copper in particular,
these temperatures are sufficient to change the
occupations of the $d$-bands $\sim 2$~eV below the Fermi level.
Consequently, it is important to derive the temperature dependence
of these material parameters from electronic structure calculations
rather than free-electron like models.\cite{Lin}

Therefore to accurately predict the transient optical response of metal nanostructures,
we account for the electron-temperature dependence of the electronic heat capacity,
electron-phonon coupling factor and dielectric functions.
These properties, in turn, require accurate electron and phonon band structures
as well as electron-phonon and optical matrix elements.
We recently showed that \emph{ab initio} calculations can quantitatively predict optical response,
carrier generation and electron transport in plasmonic metals in comparison with experiment,
with no empirical parameters.\cite{PhononAssistedDecay}
In this article, we calculate $C_e(T_e)$, $G(T_e)$ and the temperature
and frequency-dependent dielectric function, $\epsilon(\omega,T_e)$ from first principles.
These calculations implicitly include electronic-structure effects in the
density of states and electron-phonon interaction matrix elements,
and implicitly account for processes such as Umklapp scattering.
We show substantial differences between our predictions 
and those from simplified models due to the energy dependence of the
electron-phonon matrix elements, especially at high electron temperatures.

The paper is organized as follows. We start with the theoretical background
and computational methods used in the calculations of the electron
heat capacity, phonon coupling and temperature dependent
dielectric function of plasmonic materials (Section~\ref{sec:CompDetails}).
In Section~\ref{sec:DOS}, we show calculations of the electron heat capacity and its
dependence on the electron temperature due to the electronic density of states.
Analogously, section~\ref{sec:PhononDOS} presents the lattice-temperature dependence of the
lattice heat capacity due to the phonon density of states.
Next, in Section~\ref{sec:GePh} we show a key result of the paper:
temperature dependence of the electron-phonon coupling strength
accounting for energy dependence of the electron-phonon matrix elements.
Finally, section~\ref{sec:Dielectric} presents the temperature and frequency dependence
of the dielectric function, including direct (interband),
phonon-assisted and Drude intraband contributions.
Section~\ref{sec:Conclusions} summarizes our results and discusses their application
to plasmonic nanostructures in various experimental regimes.

\section{Theory and Results}

\subsection{Computational details}
\label{sec:CompDetails}

We perform density-functional theory (DFT) calculations of the electronic states,
phonons, electron-phonon and optical matrix elements,
and several derived quantities based on these properties,
for four plasmonic metals, aluminum, copper, silver and gold.
We use the open-source plane-wave density-functional software named `JDFTx'\cite{JDFTx}
to perform fully relativistic (spinorial) band structure calculations using
norm-conserving pseudopotentials at a kinetic energy cutoff of 30~Hartrees,
and the `PBEsol' exchange-correlation functional (Perdew-Burke-Ernzerhof
functional reparametrized for solids)\cite{PBEsol} with a
localized `+$U$' correction\cite{DFTplusU} for the $d$-bands in the noble metals.
Ref.~\citenum{NatCom} shows that this method produces accurate
electronic band structures in agreement with angle-resolved
photoemission (ARPES) measurements within 0.1~eV.

We calculate phonon energies and electron-phonon matrix elements using
perturbations on a $4 \times 4 \times 4$ supercell. 
In our calculations, these matrix elements implicitly 
include Umklapp-like processes.
We then convert the electron and phonon Hamiltonians to
a maximally-localized Wannier function basis,\cite{MLWFmetal}
with $12^3$ $k$-points in the Brillouin zone for electrons.
Specifically, we employ 24 Wannier centers for aluminum
and 46 spinorial centers for the noble metals which reproduces
the density functional theory (DFT)
 band structure exactly to at least 50~eV above the Fermi level.

Using this Wannier representation, we interpolate the electron, phonon
and electron-phonon interaction Hamiltonians to arbitrary wave-vectors
and perform dense Monte Carlo sampling for accurately evaluating
the Brillouin zone integrals for each derived property below.
This dense Brillouin zone sampling is necessary because of
the large disparity in the energy scales of electrons and phonons,
and directly calculating DFT phonon properties on dense $k$-point
grids is computationally expensive and impractical.
See Ref.~\citenum{PhononAssistedDecay} for further details
on the calculation protocol and benchmarks
of the accuracy of the electron-phonon coupling
(eg. resistivity within 5\% for all four metals).

\subsection{Electronic density of states and heat capacity}
\label{sec:DOS}

The electronic density of states (DOS) per unit volume
\begin{equation}
g(\varepsilon) = \int\sub{BZ}\frac{\D\vec{k}}{(2\pi)^3} \sum_n \delta(\varepsilon-\varepsilon_{\vec{k}n}),
\end{equation}
where $\varepsilon_{\vec{k}n}$ are energies of quasiparticles with band index $n$
and wave-vector $\vec{k}$ in the Brillouin zone BZ,
directly determines the electronic heat capacity and
is an important factor in the electron-phonon coupling
and dielectric response of hot electrons.
Above, the band index $n$ implicitly counts spinorial orbitals
in our relativistic calculations, and hence we omit the
explicit spin degeneracy factor.

\begin{figure}
\narrowfig{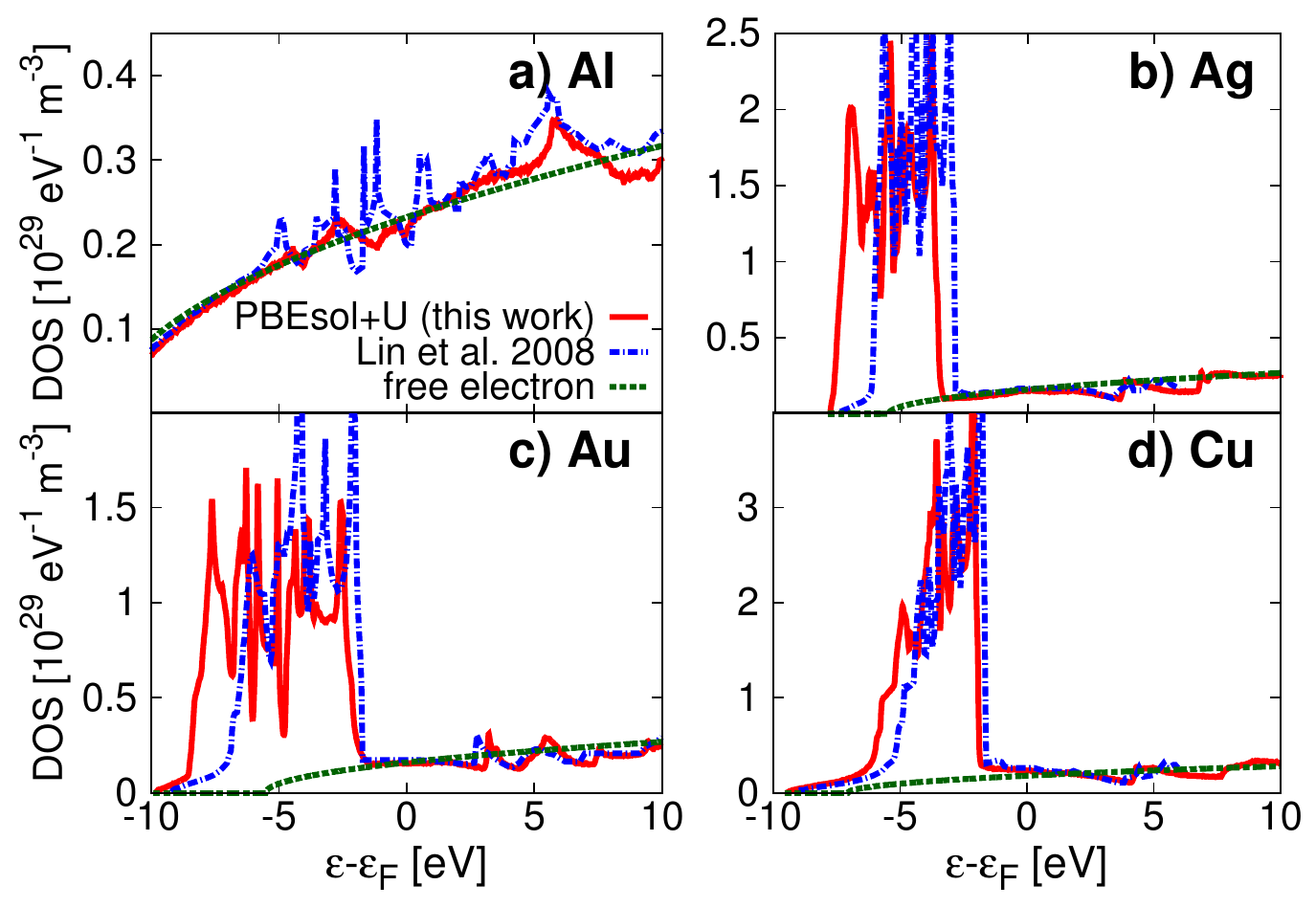}
\caption{
Comparison of electronic density of states of
for (a) Al, (b) Ag, (c) Au and (d) Cu
from our relativistic PBEsol+$U$ calculations,
previous semi-local PBE DFT calculations\cite{Lin}
(less accurate band structure), and a free electron model.
\label{fig:DOS}}
\end{figure}

Figure \ref{fig:DOS} compares the DOS predicted
by our relativistic PBEsol+$U$ method with a previous non-relativistic
semi-local estimate\cite{Lin} using the PBE (Perdew-Burke-Ernzerhof) functional,\cite{PBE}
as well as a free electron model $\varepsilon_{\vec{k}} = \frac{\hbar^2 k^2}{2m_e}$ for which
$g(\varepsilon) = \frac{\sqrt{\varepsilon}}{2\pi^2}\left(\frac{2m_e}{\hbar^2}\right)^{3/2}$.
The free electron model is a reasonable approximation for aluminum
and the PBE and PBEsol+$U$ density-functional calculations
also agree reasonably well in this case. ($U=0$ for aluminum.)
The regular $31^3$ $k$-point grid used for Brillouin zone sampling
introduces the sharp artifacts in the DOS from Ref.~\citenum{Lin}, compared to
the much denser Monte Carlo sampling in our calculations
with 640,000 $k$-points for Au, Ag, and Cu, and 1,280,000 $k$-points for Al.

For the noble metals, the free electron model and the density functional methods
agree reasonably near the Fermi level, but differ significantly $\sim 2$~eV
below the Fermi level where $d$-bands contribute.
The free electron models ignore the $d$-bands entirely, whereas
the semi-local PBE calculations predict $d$-bands that are narrower
and closer to the Fermi level than the PBEsol+$U$ predictions.
The $U$ correction\cite{DFTplusU} accounts for self-interaction errors
in semi-local DFT and positions the $d$-bands in agreement with
ARPES measurements (to within $\sim 0.1$~eV).\cite{NatCom}
Additionally, the DOS in the non-relativistic PBE calculations
strongly peaks at the top of the $d$-bands (closest to the Fermi level),
whereas the DOS in our relativistic calculations is comparatively balanced
between the top and middle of the $d$-bands due to
strong spin-orbit splitting, particularly for gold.
Below, we find that these inaccuracies in the DOS due to
electronic structure methods previously employed for studying hot electrons
propagates to the predicted electronic heat capacity and electron-phonon coupling.

\begin{figure}
\narrowfig{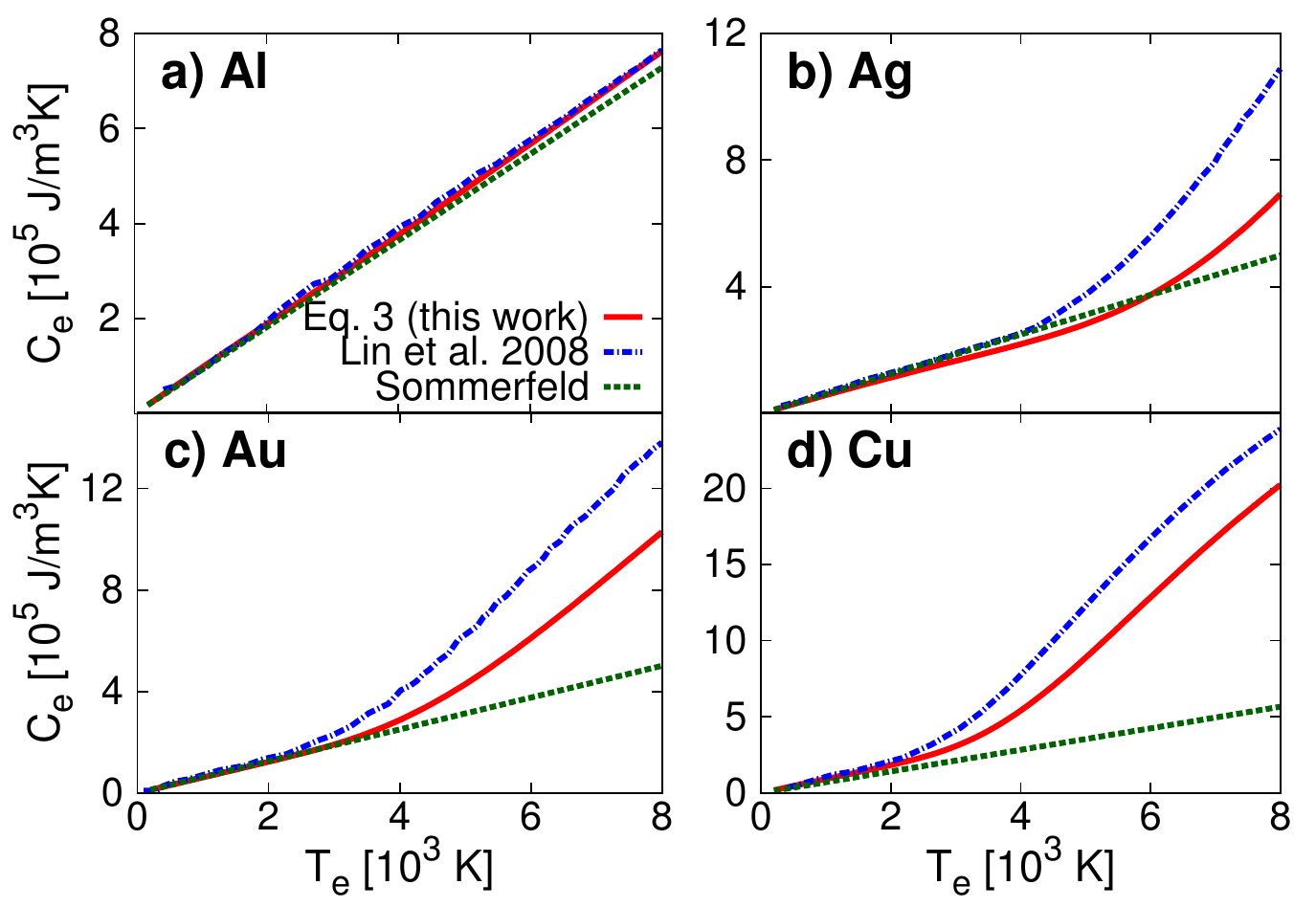}
\caption{
Comparison of the electronic heat capacity as a function
of electron temperature,  $C_e(T_e)$, for (a) Al, (b) Ag,
(c) Au and (d) Cu, corresponding to the three electronic
density-of-states predictions shown  in Figure~\ref{fig:DOS}.
The free electron Sommerfeld model underestimates $C_e$ for
noble metals at high $T_e$ because it neglects $d$-band contributions,
whereas previous DFT calculations\cite{Lin} overestimate
it because their $d$-bands are too close to the Fermi level.
\label{fig:Ce}}
\end{figure}

The electronic heat capacity, defined as the derivative of the electronic energy
per unit volume with respect to the electronic temperature ($T_e$),
can be related to the DOS as
\begin{equation}
C_e(T_e) = \int_{-\infty}^{\infty} \D\varepsilon g(\varepsilon) \varepsilon \frac{\partial f(\varepsilon,T_e)}{\partial T_e},
\label{eqn:Ce}
\end{equation}
where $f(\epsilon,T_e)$ is the Fermi distribution function.
The term $\partial f/\partial T_e$ is sharply peaked
at the Fermi energy $\varepsilon_F$ with a width $\sim k_B T_e$,
and therefore the heat capacity depends only on electronic
states within a few $k_B T_e$ of the Fermi level.
For the free electron model, Taylor expanding $g(\varepsilon)$
around $\varepsilon_F$ and analytically integrating (\ref{eqn:Ce})
yields the Sommerfeld model $C_e(T_e) = \frac{\pi^2 n_e k_B^2}{2\varepsilon_F} T_e$,
which is valid for $T_e \ll T_F$ ($\sim 10^5$~K).
Above, $n_e = 3\pi^2 k_F^3$, $\varepsilon_F = \frac{\hbar^2 k_F^2}{2m_e}$
and $k_F$ are respectively the number density, Fermi energy
and Fermi wave-vector of the free electron model.

At temperatures $T_e \ll T_F$, the electronic heat capacities
are much smaller than the lattice heat capacities,\cite{Link,Giri,Ashcroft}
which makes it possible for laser pulses to increase $T_e$
by $10^3 - 10^4$~Kelvin, while $T_l$ remains relatively constant.\cite{Hartland,Voisin,Hodak}
Figure \ref{fig:Ce} compares $C_e(T_e)$ from the free-electron Sommerfeld model
with predictions of (\ref{eqn:Ce}) using DOS from PBE and PBEsol+$U$ calculations.
The free-electron Sommerfeld model is accurate at low temperatures
(up to $\sim 2000$~K) for all four metals.

With increasing $T_e$, $\partial f/\partial T_e$ in (\ref{eqn:Ce})
is non-zero increasingly further away from the Fermi energy,
so that deviations from the free electron DOS eventually become important.
For aluminum, the DOS remains free-electron-like over a wide energy range
and the Sommerfeld model remains valid throughout.
For the noble metals, the increase in DOS due to $d$-bands causes
a dramatic increase in $C_e(T_e)$ once $T_e$ is high enough that
$\partial f/\partial T_e$ becomes non-zero in that energy range.
Copper and gold have shallower $d$-bands and deviate at lower
temperatures compared to silver. Additionally, the $d$-bands are
too close to the Fermi level in the semilocal PBE calculations of Ref.~\citenum{Lin}
which results in an overestimation of $C_e(T_e)$ compared to our predictions
based on the more accurate relativistic PBEsol+$U$ method.

\subsection{Phonon density of states and lattice heat capacity}
\label{sec:PhononDOS}

Similarly, the phonon DOS per unit volume
\begin{equation}
D(\varepsilon) = \int\sub{BZ}\frac{\D\vec{q}}{(2\pi)^3} \sum_\alpha \delta(\varepsilon-\hbar\omega_{\vec{q}\alpha}),
\label{eqn:phononDOS}
\end{equation}
where $\hbar\omega_{\vec{q}\alpha}$ are energies of phonons
with polarization index $\alpha$ and wave-vector $\vec{q}$,
directly determines the lattice heat capacity,
\begin{equation}
C_l(T_l) = \int_{0}^{\infty} \D\varepsilon D(\varepsilon) \varepsilon \frac{\partial n(\varepsilon,T_l)}{\partial T_l},
\label{eqn:Cl}
\end{equation}
where $n(\varepsilon,T_l)$ is the Bose occupation factor.

\begin{figure}
\narrowfig{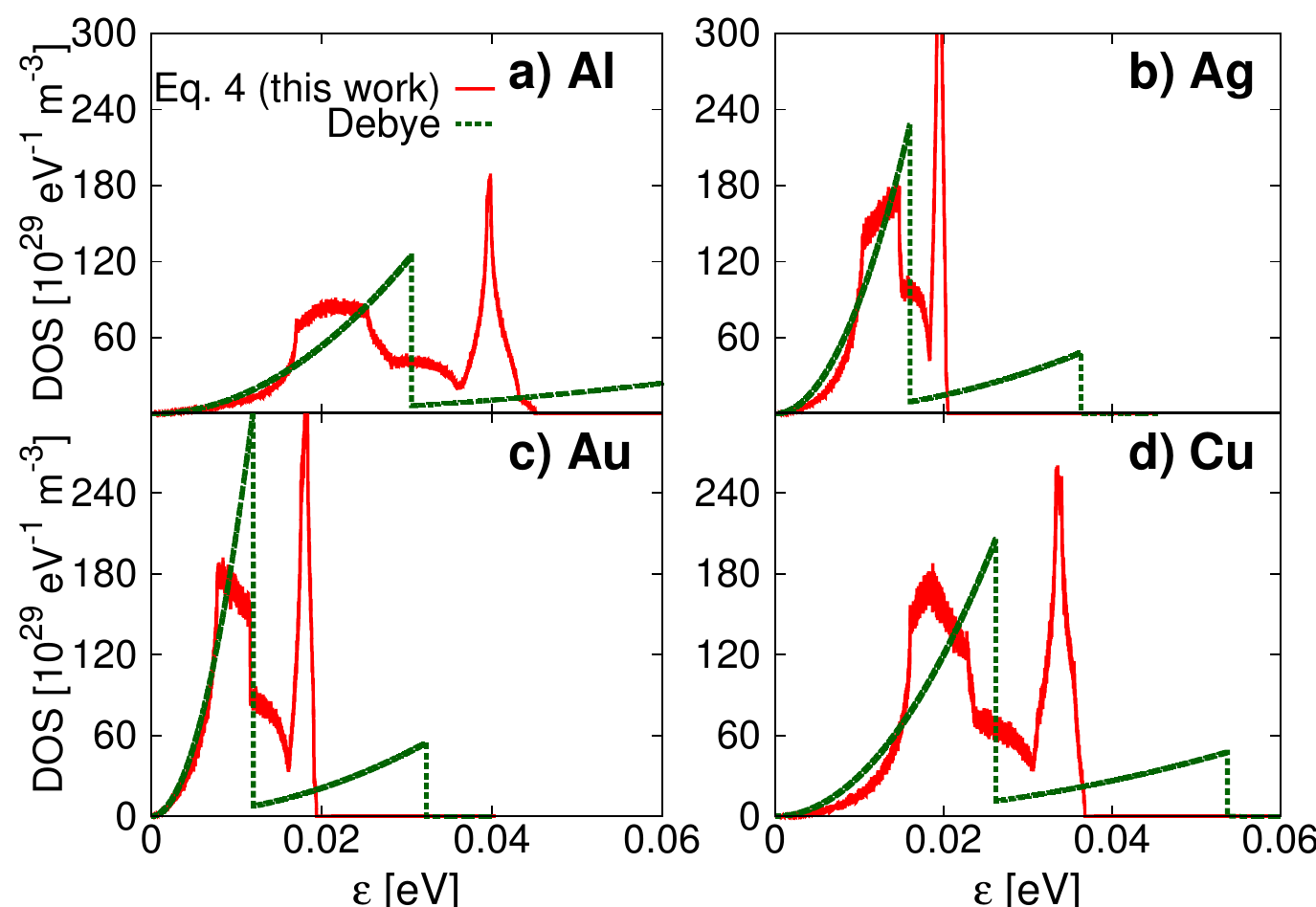}
\caption{Comparison of DFT-calculated phonon density of states
and the Debye model for (a) Al, (b) Ag, (c) Au and (d) Cu.
\label{fig:phononDOS}}
\end{figure}

Within the Debye model, the phonon energies are approximated by an
isotropic linear dispersion relation $\omega_{\vec{q}\alpha} = v_\alpha q$
up to a maximum Debye wave vector $q_D$ chosen to
conserve the number of phonon modes per unit volume.
This model yields the analytical phonon DOS,
$D(\varepsilon) = \frac{\varepsilon^2}{(2\pi^2)} \sum_\alpha \theta(\hbar q_D v_\alpha - \varepsilon)/(\hbar v_\alpha)^3$,
where $v_\alpha = \{ v_L, v_T, v_T \}$ are the speeds of sound
for the one longitudinal and two degenerate transverse phonon modes
of the face-centered cubic metals considered here.\cite{CRChandbook}

\begin{figure}
\narrowfig{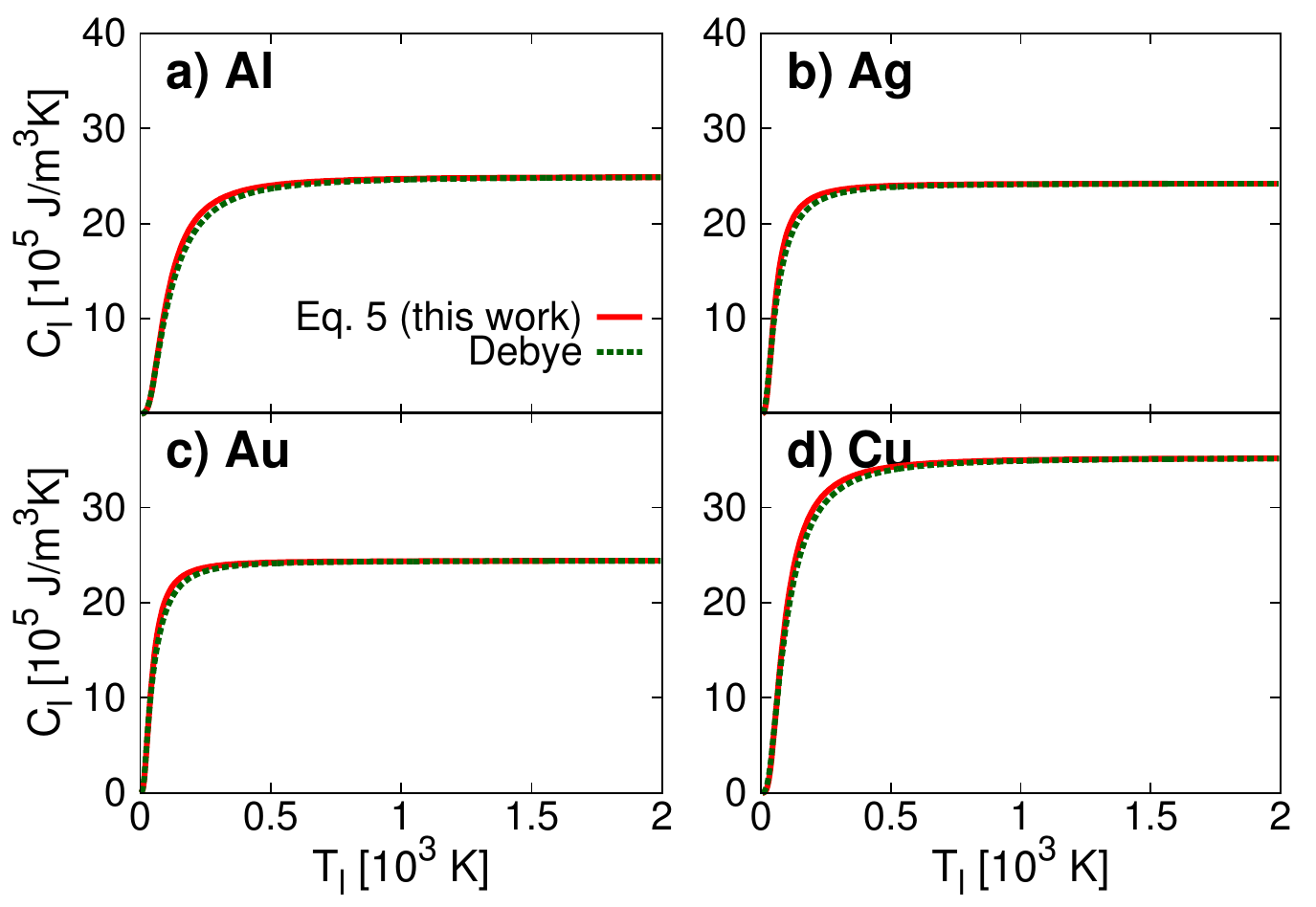}
\caption{Comparison of DFT and Debye model predictions of the
lattice heat capacity as a function of lattice temperature, $C_l(T_l)$,
for (a) Al, (b) Ag, (c) Au and (d) Cu.
Despite large differences in the density of states (Figure~\ref{fig:phononDOS}),
the predicted lattice heat capacities of the two models agree within 10\%.
\label{fig:Cl}}
\end{figure}

Figure~\ref{fig:phononDOS} compares the DFT-calculated phonon DOS
with the Debye model predictions, and shows that the Debye model
is a good approximation for the DOS only up to 0.01~eV.
However, Figure~\ref{fig:Cl} shows that the corresponding predictions
for the lattice heat capacities are very similar, rapidly approaching
the equipartition theorem prediction of $C_l = 3k_B/\Omega$ at high
temperatures, which is insensitive to details in the phonon DOS.
In fact, the largest deviations of the Debye model are below 100~K
and less than 10~\% from the direct calculations for all four metals.
We therefore find that a simple model of the phonons is adequate
for predicting the lattice heat capacity, in contrast to the remaining
quantities we consider below which are highly sensitive to
details of the phonons and their coupling to the electrons.

\subsection{Electron-phonon coupling}
\label{sec:GePh}

In Section~\ref{sec:DOS} we have shown that the electronic heat capacity,
which determines the initial temperature that the hot electrons
equilibrate to, is sensitive to electronic structure especially
in noble metals at high $T_e$ where $d$-bands contribute.
Now we analyze the electron-phonon coupling which determines
the subsequent thermalization of the hot electrons with the lattice.
We show that details in the electron-phonon matrix elements calculated using DFT
also play a significant role, in addition to the electronic band structure,
and compare previous semi-empirical estimates of the $T_e$-dependent
phonon coupling to our direct calculations.

The rate of energy transfer from electrons at temperature $T_e$
to the lattice (phonons) at temperature $T_l$ per unit volume
is given by Fermi's golden rule as
\begin{flalign}
\frac{\D E}{\D t} &\equiv G(T_e) (T_e - T_l)\\
&= \frac{2\pi}{\hbar} \int\sub{BZ}\frac{\Omega \D\vec{k}\D\vec{k}'}{(2\pi)^6} \sum_{nn'\alpha}
\delta(b - \hbar\omega_{\vec{k}'-\vec{k},\alpha}) \nonumber\\
&\hspace{2ex}\times
	\hbar\omega_{\vec{k}'-\vec{k},\alpha}
	\left|g^{\vec{k}'-\vec{k},\alpha}_{\vec{k}'n',\vec{k}n} \right|^2
	S_{T_e,T_l}(\varepsilon_{\vec{k}n}, \varepsilon_{\vec{k}'n'}, \hbar\omega_{\vec{k}'-\vec{k},\alpha})
\nonumber
\end{flalign}
with
\begin{multline}
S_{T_e,T_l}(\varepsilon,\varepsilon',\hbar\omega\sub{ph})
	\equiv f(\varepsilon,T_e) n(\hbar\omega\sub{ph},T_l) (1-f(\varepsilon',T_e)) \\
	 - (1-f(\varepsilon,T_e)) (1+n(\hbar\omega\sub{ph},T_l)) f(\varepsilon',T_e).
\end{multline}
Here, $\Omega$ is the unit cell volume, $\hbar\omega_{\vec{q}\alpha}$
is the energy of a phonon with wave-vector $\vec{q}=\vec{k}'-\vec{k}$ and
polarization index $\alpha$, and $g^{\vec{k}'-\vec{k},\alpha}_{\vec{k}'n',\vec{k}n}$
is the electron-phonon matrix element coupling this phonon
to electronic states indexed by $\vec{k}n$ and $\vec{k}'n'$.

Above, $S$ is the difference between the product of occupation factors
for the forward and reverse directions of the electron-phonon
scattering process $\vec{k}n + \vec{q}\alpha \rightarrow \vec{k}'n'$,
with $f(\varepsilon,T_e)$ and $n(\hbar\omega,T_l)$ being the Fermi
and Bose distribution function for the electrons and phonons respectively.
Using the fact that $S_{T_e,T_e} = 0$ for an energy-conserving process
$\varepsilon + \hbar\omega\sub{ph} = \varepsilon'$ by detailed balance,
we can write the electron-phonon coupling coefficient as
\begin{multline}
G(T_e) = \frac{2\pi}{\hbar} \int\sub{BZ}\frac{\Omega \D\vec{k}\D\vec{k}'}{(2\pi)^6} \sum_{nn'\alpha}
\delta(\varepsilon_{\vec{k}'n'} - \varepsilon_{\vec{k}n} - \hbar\omega_{\vec{k}'-\vec{k},\alpha}) \\
\times
	\hbar\omega_{\vec{k}'-\vec{k},\alpha}
	\left|g^{\vec{k}'-\vec{k},\alpha}_{\vec{k}'n',\vec{k}n} \right|^2
	(f(\varepsilon_{\vec{k}n},T_e) - f(\varepsilon_{\vec{k}'n'},T_e))\\
\times
	\frac{n(\hbar\omega_{\vec{k}'-\vec{k},\alpha},T_e) - n(\hbar\omega_{\vec{k}'-\vec{k},\alpha},T_l)}{T_e - T_l}
\label{eqn:G}
\end{multline}
This general form for \emph{DFT-based} electronic and phononic states
is analogous to previous single-band / free electron theories of the
electron-phonon coupling coefficient, see for example the derivation by Allen et al.\cite{Allen}
Note that unlike previous empirical models, here the coupling coefficient
depends on the lattice temperature $T_l$ as well, but we omit the $T_l$ label
in $G(T_e)$ to keep the notation consistent with previous approaches,\cite{Lin}
and present results below for $T_l = 298$~K (ambient temperature).

The direct evaluation of $G(T_e)$ using (\ref{eqn:G})
requires a six-dimensional integral over electron-phonon matrix elements
from DFT with very fine $k$-point grids that can resolve
both electronic and phononic energy scales.
This is impractical without the recently-developed Wannier interpolation
and Monte Carlo sampling methods for these matrix elements,\cite{EPW,PhononAssistedDecay}
and therefore our results are the first parameter-free predictions
of $G(T_e)$, derived entirely from DFT.

Previous theoretical estimates of $G(T_e)$ are semi-empirical,
combining DFT electronic structure with empirical models for the phonon coupling.
For example, Wang et al.\cite{Wang} assume that the electron-phonon matrix elements
averaged over scattering angles is independent of energy and that the
phonon energies are smaller than $k_B T_e$, and then approximate
the electron-phonon coupling coefficient as
\begin{equation}
G(T_e) \approx \frac{\pi k_B}{\hbar g(\varepsilon_F)} \lambda\langle(\hbar\omega)^2\rangle
	\int_{-\infty}^\infty \D\varepsilon g^2(\varepsilon) \frac{-\partial f(\varepsilon,T_e)}{\partial \varepsilon},
\label{eqn:GLin}
\end{equation}
where $\lambda$ is the electron-phonon mass enhancement parameter and $\langle(\hbar\omega)^2\rangle$
is the second moment of the phonon spectrum.\cite{Grimvall,McMillan,Lin}
Lin et al.\cite{Lin} treat $\lambda\langle (\hbar\omega)^2 \rangle$ as an empirical parameter
calibrated to experimental $G(T_e)$ at low $T_e$ obtained from thermoreflectance measurements,
and extrapolate it to higher $T_e$ using (\ref{eqn:GLin}).
See Refs.~\citenum{Wang} and \citenum{Lin} for more details.

For clarity, we motivate here a simpler derivation of an expression of
the form of (\ref{eqn:GLin}) from the general form (\ref{eqn:G}).
First, making the approximation $\hbar\omega_{\vec{q}\alpha}\ll T_e$
(which is reasonably valid for $T_e$ above room temperature)
allows us to approximate the difference between the electron occupation factors
in the second line of (\ref{eqn:G}) by $\hbar\omega_{\vec{q}\alpha}
\partial f/\partial \varepsilon$ (using energy conservation).
Additionally, for $T_e \gg T_l$, the third line of (\ref{eqn:G})
simplifies to $k_B/(\hbar\omega_{\vec{k}'-\vec{k},\alpha})$.
With no other approximations, we can then rearrange (\ref{eqn:G})
to collect contributions by initial electron energy,
\begin{equation}
G(T_e) \approx \frac{\pi k_B}{\hbar g(\varepsilon_F)} \int_{-\infty}^\infty \D\varepsilon
	h(\varepsilon) g^2(\varepsilon) \frac{-\partial f(\varepsilon,T_e)}{\partial\varepsilon} 
\end{equation}
with
\begin{multline}
h(\varepsilon) \equiv \frac{2g(\varepsilon_F)}{g^2(\varepsilon)}
	\int\sub{BZ}\frac{\Omega \D\vec{k}\D\vec{k}'}{(2\pi)^6} \sum_{nn'\alpha}
\delta(\varepsilon - \varepsilon_{\vec{k}n})\\
\times
	\delta(\varepsilon_{\vec{k}'n'} - \varepsilon_{\vec{k}n} - \hbar\omega_{\vec{k}'-\vec{k},\alpha}) 
	\hbar\omega_{\vec{k}'-\vec{k},\alpha}
	\left|g^{\vec{k}'-\vec{k},\alpha}_{\vec{k}'n',\vec{k}n} \right|^2.
\label{eqn:h}
\end{multline}
Therefore, the primary approximation in previous semi-empirical estimates\cite{Lin,Wang}
is the replacement of $h(\varepsilon)$ by an energy-independent constant
$\lambda\langle (\hbar\omega)^2 \rangle$, used as an empirical parameter.

\begin{figure}
\narrowfig{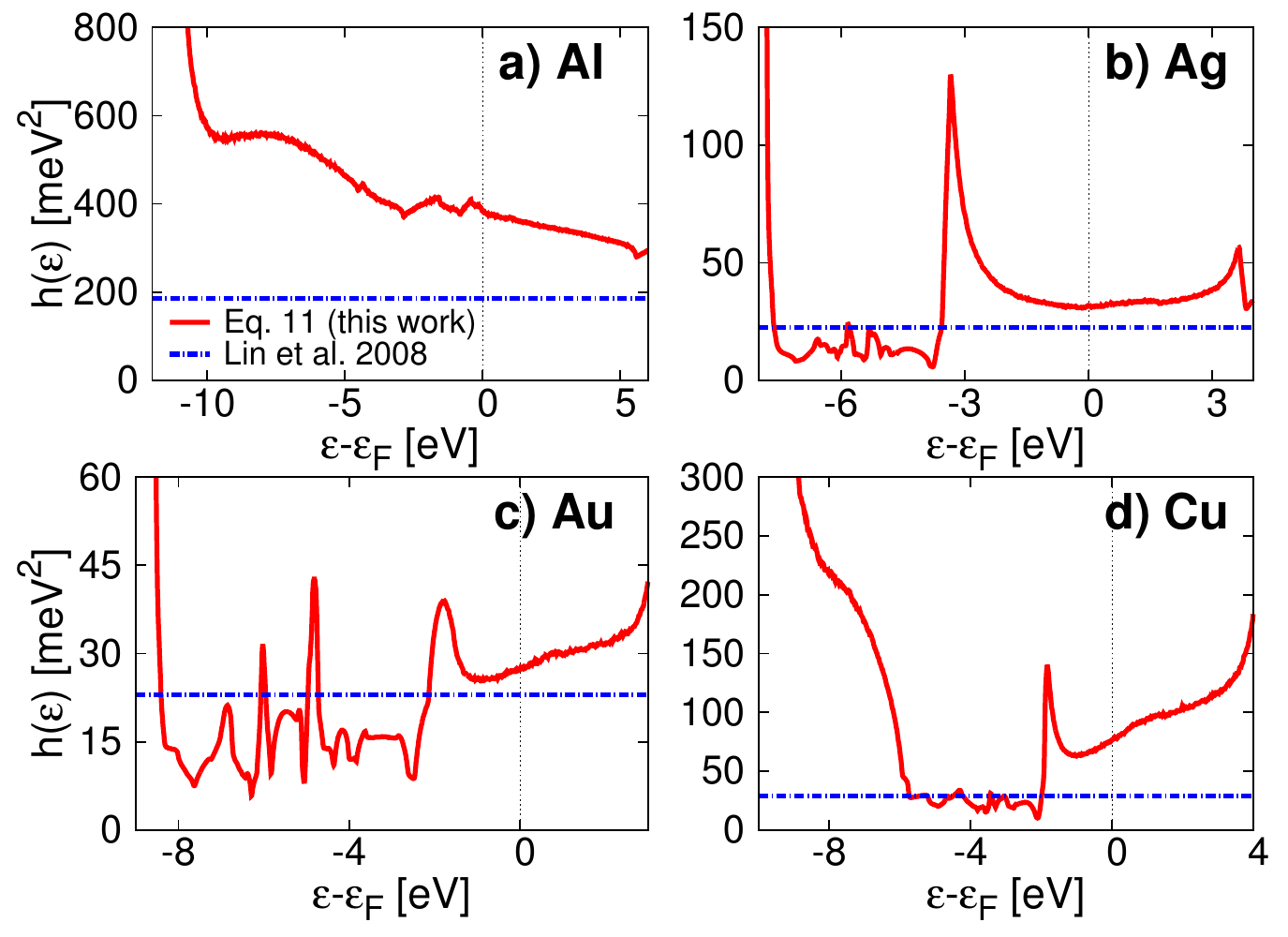}
\caption{
Energy-resolved electron-phonon coupling strength $h(\varepsilon)$,
defined by (\ref{eqn:h}), for (a) Al, (b) Ag, (c) Au, (d) Cu.
For the noble metals, $h(\epsilon_F)$ is substantially larger
than its value in the $d$-bands, which causes previous semi-empirical
estimates\cite{Lin} using a constant $h(\varepsilon)$ to overestimate
the electron-phonon coupling ($G(T_e)$) at $T_e \gtrsim 3000$~K,
as shown in Fig.~\ref{fig:G}.
\label{fig:h}}
\end{figure}

\begin{figure}
\narrowfig{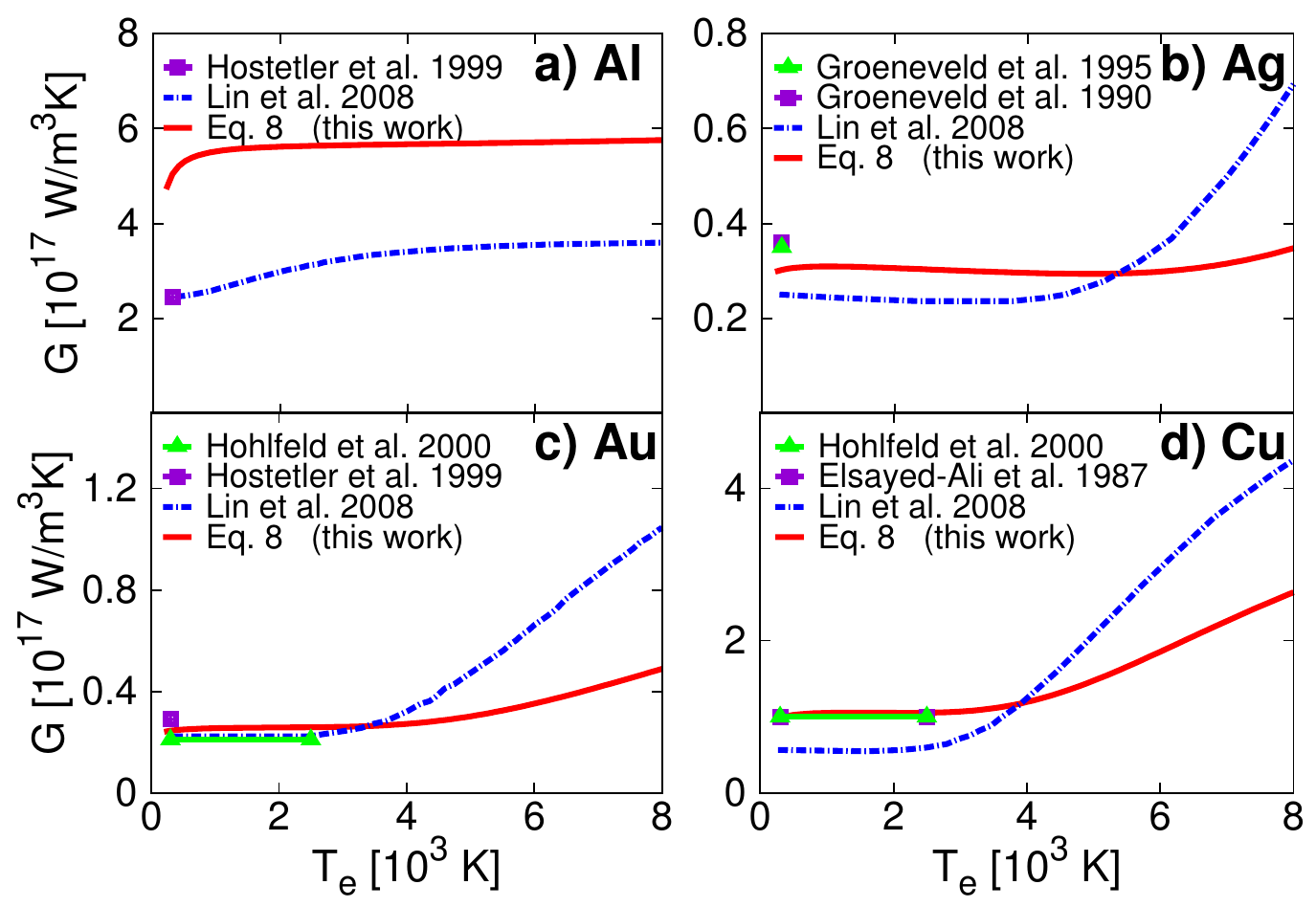}
\caption{
Comparison of predictions of the electron-phonon
coupling strength as a function of electron temperature, $G(T_e)$,
for (a) Al, (b) Ag, (c) Au and (d) Cu, with experimental measurements where
available.\cite{ Hostetler,Hohlfeld,Elsayed1987,Groeneveld1990,Groeneveld1995}
The DFT-based semi-empirical predictions of Lin et al.\cite{Lin} overestimate
the coupling for noble metals at high temperatures because they assume
an energy-independent electron-phonon coupling strength (Figure~\ref{fig:h})
and neglect the weaker phonon coupling of $d$-bands compared to the conduction band.
The experimental results (and hence the semi-empirical predictions) for aluminum
underestimate electron-phonon coupling because they include the effect of
competing electron-electron thermalization which happens on the same time scale.
\label{fig:G}}
\end{figure}

Fig.~\ref{fig:h} compares our calculations of this
energy-resolved electron-phonon coupling strength, $h(\varepsilon)$,
with previous empirical estimates of $\lambda\langle (\hbar\omega)^2 \rangle$,
and Fig.~\ref{fig:G} compares the resulting temperature dependence
of the electron-phonon coupling, $G(T_e)$, from
(\ref{eqn:G}) and semi-empirical methods(\ref{eqn:GLin}).
For noble metals, $G(T_e)$ increases sharply beyond $T_e\sim 3000$~K
because of the large density of states in the $d$-bands.
However, $h(\varepsilon)$ is smaller by a factor of $2-3$
in the $d$-bands compared to near the Fermi level.
Therefore, assuming $h(\varepsilon)$ to be an empirical constant\cite{Lin,Sun}
results in a significant overestimate of $G(T_e)$ at high $T_e$,
compared to the direct calculations.
Additionally, the shallowness of the $d$-bands in the semi-local PBE band structure
used in Ref.~\citenum{Lin} lowers the onset temperature of the increase in $G(T_e)$,
and results in further overestimation compared to our predictions.

Our predictions agree very well with the experimental
measurements of $G(T_e)$ available at lower temperatures for noble metals.\cite{Hostetler,Hohlfeld,Elsayed1987,Groeneveld1990,Groeneveld1995}
In fact, the semi-empirical calculation based on $\lambda\langle(\hbar\omega)^2\rangle$
underestimates the room temperature electron-phonon coupling for these metals;
the significant overestimation of $G(T_e)$ seen in Fig.~\ref{fig:G} is
in despite this partial cancellation of error.
This shows the importance of detailed DFT electron-phonon matrix elements
in calculating the coupling between hot electrons and the lattice.

Experimental measurements of the electron-phonon coupling
in noble metals are reliable because of the reasonably clear separation
between a fast electron-electron thermalization rise followed by
a slower electron-phonon decay in the thermoreflectance signal.
In aluminum, these time scales significantly overlap
resulting in strong non-equilibrium effects and making experimental
determination of the equilibrium electron-phonon coupling $G(T_e)$ difficult.
Consequently, the value of $G(T_e)$ for Al is not well agreed upon.\cite{Hostetler,Guo}
Using a simplified single-band free-electron-like model of the electrons,
Ref.~\citenum{Mueller:2013} estimates $G \approx \sci{2.9}{7}$~W/m\super{3}K
for thermalized electrons at 2000~K, which is 1.5x larger than
$G \approx \sci{1.9}{7}$~W/m\super{3}K for non-thermalized electrons
with the same amount of energy.
In figure~\ref{fig:G}(a), our predictions using (\ref{eqn:G})
which assumes equilibrium are 2x larger than the experimental
estimates\cite{Hostetler} which implicitly include the non-equilibrium effects.
On the other hand, the semi-empirical model of Ref.~\citenum{Lin}
assumes thermalized electrons, but fits to experimental data
that includes non-thermal effects (and matches experiment by construction).
The single-band-model non-equilibrium predictions do not match
experiment because it assumes a simple model for electron-phonon
matrix elements that ignores Umklapp processes.\cite{Mueller:2013}
Ultimately, quantitative agreement with experiments for aluminum
(for the right reasons) therefore requires an extension of our
non-empirical DFT approach (\ref{eqn:G}) to include non-equilibrium effects,
a subject of current work in our group.

\subsection{Dielectric Function}
\label{sec:Dielectric}

The final ingredient for a complete theoretical description
of ultrafast transient absorption measurements is the
temperature-dependent dielectric function of the material.
We previously showed\cite{PhononAssistedDecay} that we could predict the imaginary part
of the dielectric function $\Im\epsilon(\omega)$ of plasmonic metals in quantitative
agreement with ellipsometric measurements for a wide range of frequencies
by accounting for the three dominant contributions,
\begin{equation}
\Im\epsilon(\omega) = \frac{4\pi \sigma_0}{\omega(1+\omega^2\tau^2)}
	+ \Im\epsilon\sub{direct}(\omega) + \Im\epsilon\sub{phonon}(\omega).
\label{eqn:ImEpsTot}
\end{equation}
We briefly summarize the calculation of these contributions 
and focus on their electron temperature dependence below;
see Ref.~\citenum{PhononAssistedDecay} for a detailed description.

The first term of (\ref{eqn:ImEpsTot}) accounts for the Drude response of the metal
due to free carriers near the Fermi level, with the zero-frequency conductivity
$\sigma_0$ and momentum relaxation time $\tau$ calculated using
the linearized Boltzmann equation with collision integrals based on DFT.\cite{PhononAssistedDecay}
The second and third terms of (\ref{eqn:ImEpsTot}),
\begin{widetext}
\begin{equation}
\Im\epsilon\sub{direct}(\omega)
= \frac{4\pi^2 e^2}{m_e^2\omega^2} \int\sub{BZ} \frac{d\vec{k}}{(2\pi)^3} \sum_{n'n}
	(f_{\vec{k}n} - f_{\vec{k}n'}) \delta(\varepsilon_{\vec{k}n'} - \varepsilon_{\vec{k}n} - \hbar\omega)
	\left| \hat{\lambda} \cdot \langle\vec{p}\rangle^{\vec{k}}_{n'n} \right|^2,
\textrm{ and}
\label{eqn:ImEpsDirect}
\end{equation}
\begin{multline}
\Im\epsilon\sub{phonon}(\omega)
= \frac{4\pi^2 e^2}{m_e^2\omega^2}
	\int\sub{BZ} \frac{d\vec{k}'d\vec{k}}{(2\pi)^6} \sum_{n'n\alpha\pm}
	(f_{\vec{k}n} - f_{\vec{k}'n'}) \left( n_{\vec{k}'-\vec{k},\alpha} + \frac{1}{2} \mp \frac{1}{2}\right)
	\delta(\varepsilon_{\vec{k}'n'} - \varepsilon_{\vec{k}n} - \hbar\omega \mp \hbar\omega_{\vec{k}'-\vec{k},\alpha}) \\
\times \left| \hat{\lambda} \cdot \sum_{n_1} \left(
	\frac{ g^{\vec{k}'-\vec{k},\alpha}_{\vec{k}'n',\vec{k}n_1} \langle\vec{p}\rangle^{\vec{k}}_{n_1n} }{ \varepsilon_{\vec{k}n_1} - \varepsilon_{\vec{k}n} - \hbar\omega + i\eta} +
	\frac{ \langle\vec{p}\rangle^{\vec{k}'}_{n'n_1} g^{\vec{k}'-\vec{k},\alpha}_{\vec{k}'n_1,\vec{k}n} }{ \varepsilon_{\vec{k}'n_1} - \varepsilon_{\vec{k}n} \mp \hbar \omega_{\vec{k}'-\vec{k},\alpha} + i\eta }
\right) \right|^2,
\label{eqn:ImEpsPhonon}
\end{multline}
\end{widetext}
capture the contributions due to direct interband excitations
and phonon-assisted intraband excitations respectively.
Here $\langle\vec{p}\rangle^{\vec{k}}_{n'n}$ are matrix elements of the
momentum operator, $\hat{\lambda}$ is the electric field direction
(results are isotropic for crystals with cubic symmetry), and all
remaining electron and phonon properties are exactly as described previously.
The energy-conserving $\delta$-functions are replaced by a Lorentzian
of width equal to the sum of initial and final electron linewidths,
because of the finite lifetime of the quasiparticles.

The dielectric function calculated using (\ref{eqn:ImEpsTot}-\ref{eqn:ImEpsPhonon})
depends on the electron temperature $T_e$ in two ways.
First, the electron occupations $f_{\vec{k}n}$ directly depend on $T_e$.
Second, the phase-space for electron-electron scattering increases with
electron temperature, which increases the Lorentzian broadening in the energy 
conserving $\delta$-functions in (\ref{eqn:ImEpsDirect}) and (\ref{eqn:ImEpsPhonon}).

We calculate electron linewidths from DFT using Fermi golden rule
calculations for electron-electron and electron-phonon scattering
at room temperature, as detailed in Ref.~\citenum{PhononAssistedDecay}.
These calculations are computationally expensive and difficult to repeat
for several electron temperatures; we instead use the linewidths
at room temperature with an analytical correction for the $T_e$ dependence.
The electron-phonon scattering rate depends on the lattice temperature,
but is approximately independent of $T_e$ because the phase space for scattering
is determined primarily by the electronic density-of-states and
electron-phonon matrix elements, which depend strongly on the
electron energies but not on the occupation factors or $T_e$.
The phase space for electron-electron scattering, on the other hand,
depends on the occupation factors and $T_e$ because an electron
at an energy far from the Fermi level can scatter with electrons
close to the Fermi level.
The variation of this phase-space with temperature is primarily
due to the change in occupation of states near the Fermi level,
and we can therefore estimate this effect in plasmonic metals
using a free electron model.

Within a free electron model, the phase-space for electron-electron
scattering grows quadratically with energy relative to the Fermi level,
resulting in scattering rates $\propto (\varepsilon-\varepsilon_F)^2$
at zero electron temperatures, as is well-known.\cite{DelFatti,eeLinewidth}
We can extend these derivations to finite electron temperature to show that
the energy and temperature-dependent electron-electron scattering rate
\begin{equation}
\tau^{-1}\sub{ee}(\varepsilon,T_e) \approx \frac{D_e}{\hbar} [ (\varepsilon - \varepsilon_F)^2 + (\pi k_B T_e)^2 ]
\label{eqn:tauTeDep}
\end{equation}
for $|\varepsilon-\varepsilon_F| \ll \varepsilon_F$ and $T_e \ll \varepsilon_F/k_B$.
Within the free electron model, the constant of proportionality
$D_e = \frac{m_e e^4}{4\pi\hbar^2(\epsilon_b^0)^2 \varepsilon_S^{3/2} \sqrt{\varepsilon_F}}
\left( \frac{\sqrt{4\varepsilon_F\varepsilon_S}}{4\varepsilon_F + \varepsilon_S}
+ \tan^{-1}\sqrt{\frac{4\varepsilon_F}{\varepsilon_S}} \right)$,
where the background dielectric constant $\epsilon_b^0$ and
the Thomas-Fermi screening energy scale $\varepsilon_S$ are
typically treated as empirical parameters.\cite{DelFatti}
Here, we extract $D_e$ by fitting (\ref{eqn:tauTeDep})
to the electron-electron scattering rates at room temperature $T_0$
calculated using DFT.\cite{PhononAssistedDecay}
The resulting fit parameters are listed in Table~\ref{tab:De}.
We then estimate the total scattering rates at other temperatures by
adding $(D_e/\hbar) (\pi k_B)^2 (T_e^2 - T_0^2)$ to the total DFT-calculated
results (including electron-phonon scattering) at $T_0$.
Note that we could have equivalently fit the DFT-calculated scattering rates
at zero temperature, but the Fermi Golden rule results at room temperature
are less noisy at finite $k$-point sampling, and moreover these rates do not
differ appreciably for electron energies more than $\sim \pi k_B T_0 \approx 0.08$~eV
away from the Fermi level anyway.

\begin{table}
\caption{
Coefficient of the temperature dependence of the electron-electron scattering rate
as given by (\ref{eqn:tauTeDep}), extracted from fits to the energy dependence of
DFT-calculated electron-electron scattering rates at room temperature.\cite{PhononAssistedDecay}
\label{tab:De}}
\begin{tabular}{lcccc}
\hline\hline
Metal & Al & Ag & Au & Cu \\
\hline
$D_e$~[eV\super{-1}] & 0.017 & 0.021 & 0.016 & 0.020\\
\hline\hline
\end{tabular}
\end{table}

\begin{figure}
\narrowfig{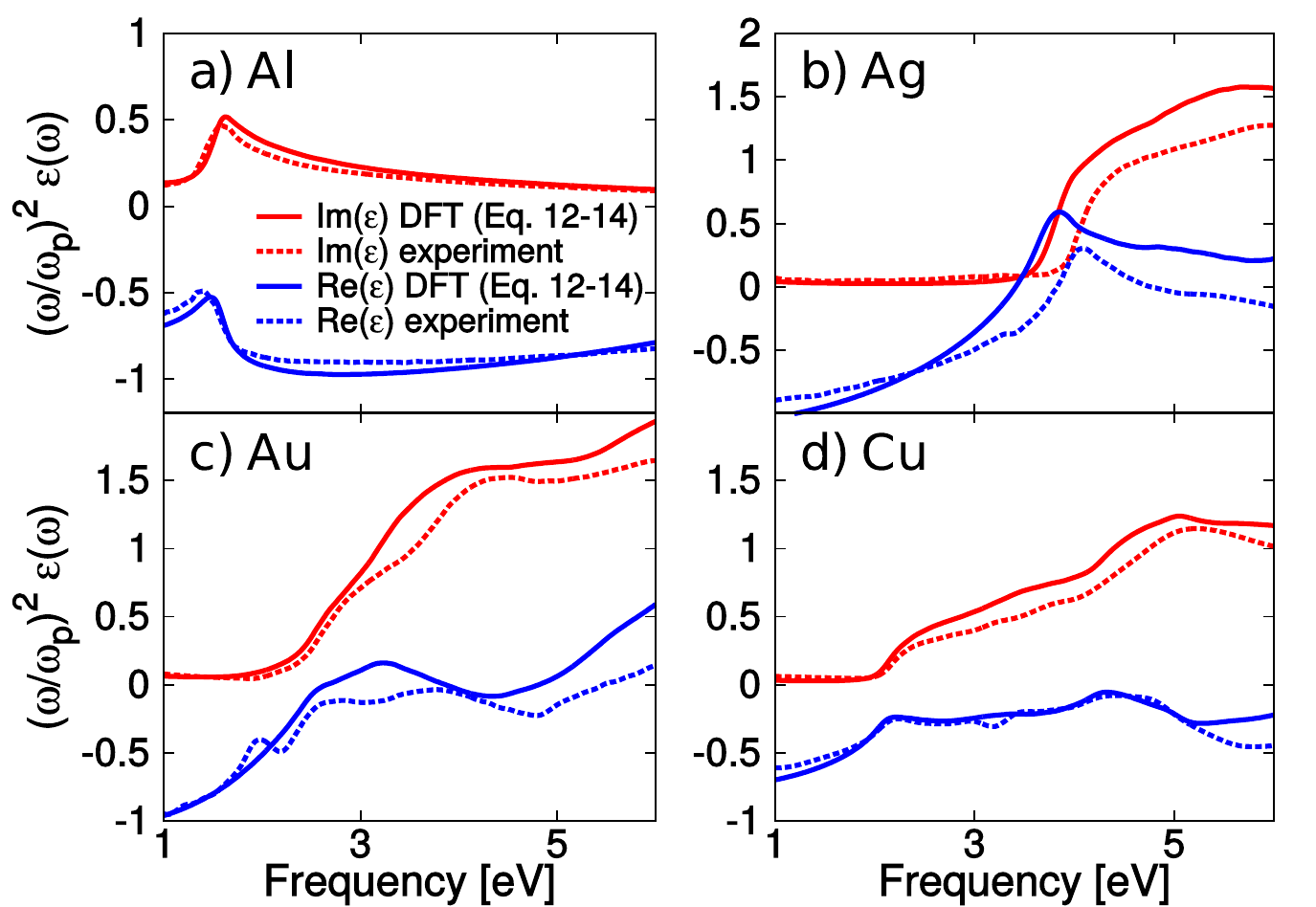}
\caption{
DFT predictions of the complex dielectric functions for (a) Al, (b) Ag, (c) Au, (d) Cu
at room temperature (300~K) compared with ellipsometry measurements.\cite{Palik1985}
The $y$-axis is scaled by $\omega^2 / \omega_p^2$ in order to represent
features at different frequencies such as the Drude pole
and the interband response on the same scale.
\label{fig:Eps300K}}
\end{figure}

Finally, we use the Kramers-Kronig relations to calculate
$\Re(\epsilon(\omega,T_e))$ from $\Im(\epsilon(\omega,T_e))$.
Figure~\ref{fig:Eps300K} compares the DFT-predicted dielectric functions
with ellipsometry measurements\cite{Palik1985} for a range
of frequencies spanning from near-infrared to ultraviolet.
Note that we scale the $y$-axis by $(\omega/\omega_p)^2$, where
$\omega_p = \sqrt{4\pi e^2 n_e/m_e}$ is the free-electron plasma frequency,
in order to display features at all frequencies on the same scale.
We find excellent agreement for aluminum within 10~\% of experiment
over the entire frequency range, including the peak around 1.6~eV
due to an interband transition.
The agreement is reasonable for noble metals with a typical error
within 20~\%, but with a larger error  $\sim50$~\% for certain features
in the interband $d\rightarrow s$ transitions due to inaccuracies
in the $d$-band positions predicted by DFT (especially for silver).
In the present work, the PBEsol+$U$ band structure is typically
accurate to $\sim 0.1$~eV,\cite{NatCom} compared to errors $\sim 1$~eV
in $d$-band positions predicted by semi-local DFT functionals\cite{Lin} and
qualitative inadequacies of free-electron-like models that ignore $d$ bands entirely.
Consequently, our chosen method has the potential to provide the most reliable predictions of
metal dielectric functions, especially for the electron temperature dependence that we discuss next.
(Empirical fits such as Drude-Lorentz models can be more accurate by construction at
one temperature,\cite{DrudeLorentzDielFunc} but do not predict temperature dependence.)

\begin{figure}
\narrowfig{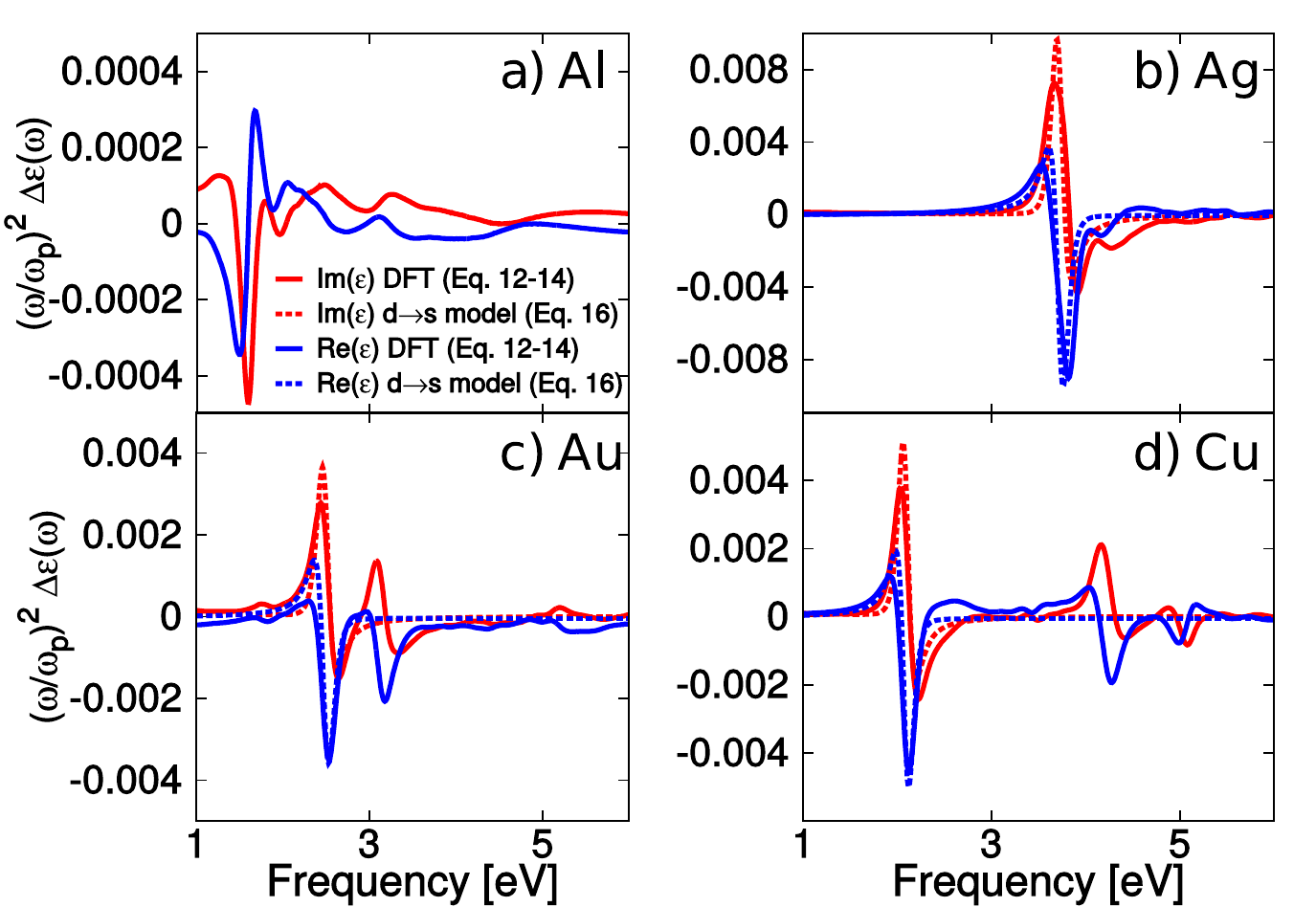}
\caption{
Change in the DFT-predicted complex dielectric function (solid lines) for
(a) Al, (b) Ag, (c) Au, (d) Cu from room temperature
(300~K) to electron temperature $T_e = 400$~K 
(with the lattice remaining at room temperature).
In comparison, the analytical $d \rightarrow s$ model (\ref{eqn:EpsChangeModel})
(dashed lines) captures essential features of the DFT results
for noble metals at lower temperatures, but misses the contributions
of broadening due to electron-electron scattering at higher temperatures.
Note that the $y$-axis is scaled as in Fig.~\ref{fig:Eps300K} for clarity.
\label{fig:EpsChange400K}}
\end{figure}

\begin{figure}
\narrowfig{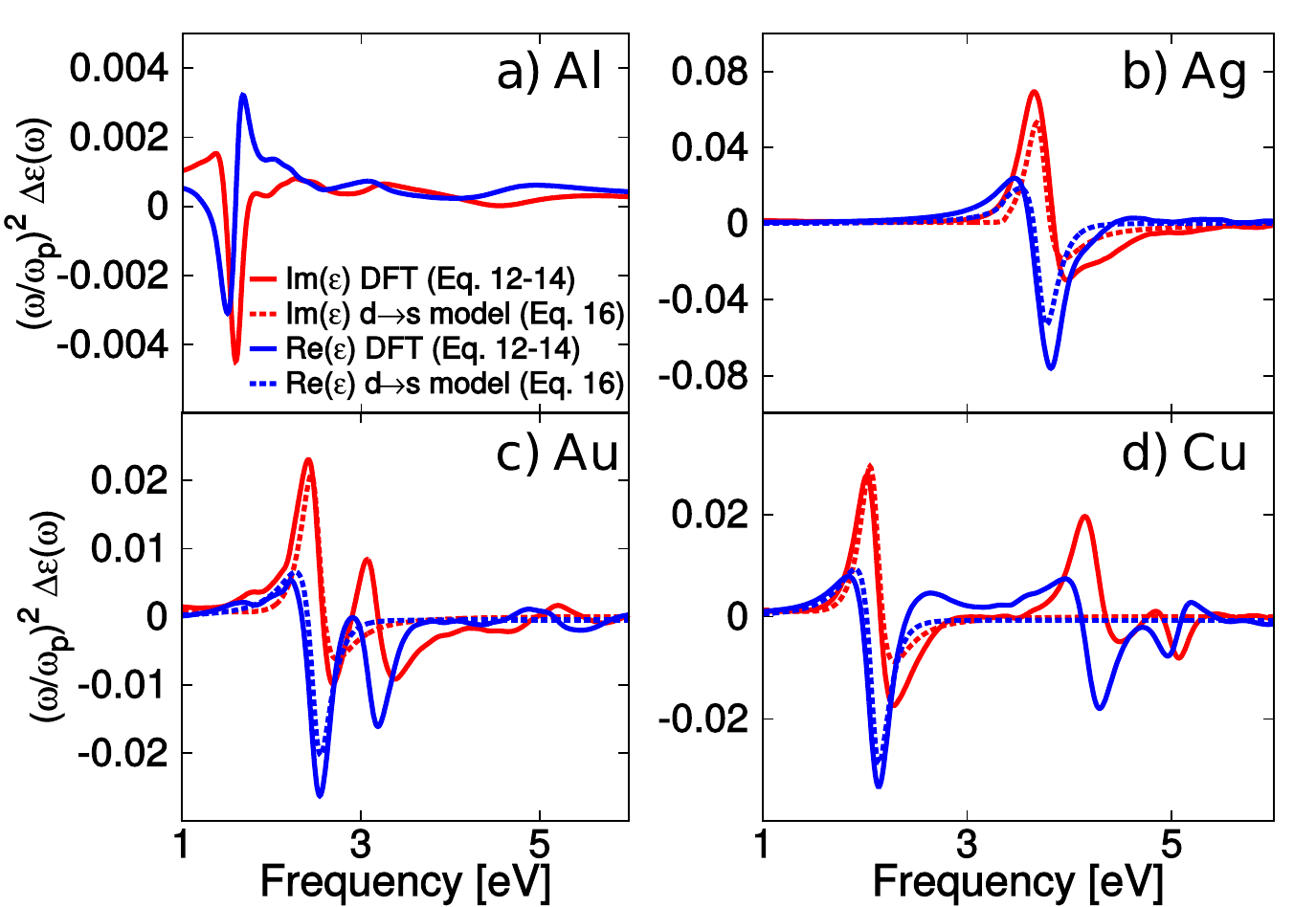}
\caption{
Change in the DFT-predicted complex dielectric function (solid lines) for
(a) Al, (b) Ag, (c) Au, (d) Cu from room temperature
(300~K) to electron temperature $T_e = 1000$~K 
(with the lattice remaining at room temperature),
compared to the analytical $d \rightarrow s$ model
(\ref{eqn:EpsChangeModel}) (dashed lines).
\label{fig:EpsChange1000K}}
\end{figure}

\begin{figure}
\narrowfig{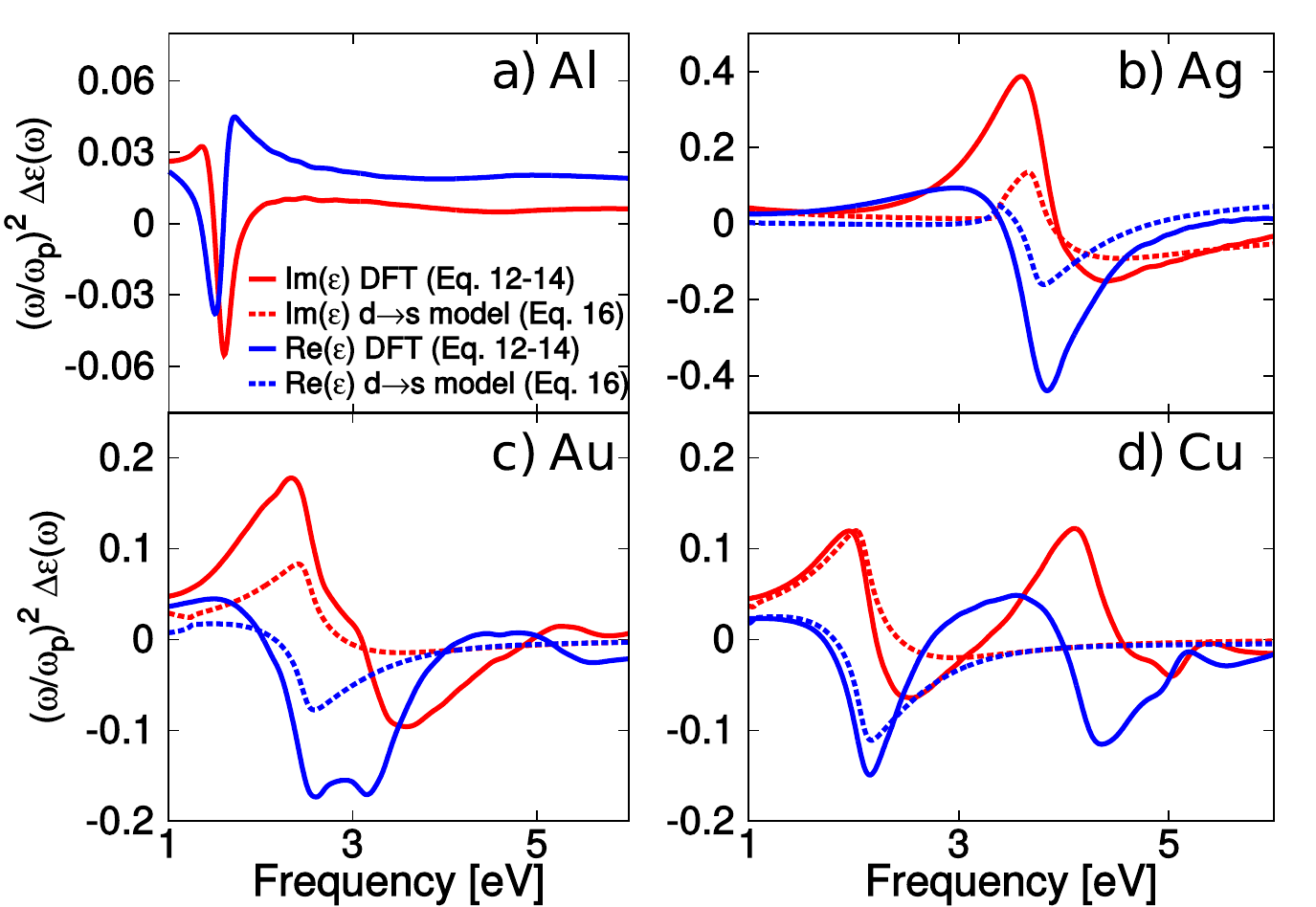}
\caption{
Change in the DFT-predicted complex dielectric function (solid lines) for
(a) Al, (b) Ag, (c) Au, (d) Cu from room temperature
(300~K) to electron temperature $T_e = 5000$~K 
(with the lattice remaining at room temperature),
compared to the analytical $d \rightarrow s$ model
(\ref{eqn:EpsChangeModel}) (dashed lines).
\label{fig:EpsChange5000K}}
\end{figure}

Figures~\ref{fig:EpsChange400K}, \ref{fig:EpsChange1000K} and
\ref{fig:EpsChange5000K} show the change of the DFT-calculated
complex dielectric function (solid lines) upon increasing the electron temperature $T_e$
from room temperature to 400~K, 1000~K and 5000~K respectively,
while the lattice remains at room temperature.\cite{NoteSI}
For all four metals, the response from infrared to ultraviolet frequencies
is dominated by `sharp' features due to interband transitions 
that broaden with increasing temperature.
In the remainder of this section, we analyze these sharp interband features
in greater detail using a simpler analytic model of the ($d \rightarrow s$) transitions
(shown in dashed lines in the aforementioned figures).

\begin{figure}
\narrowfig{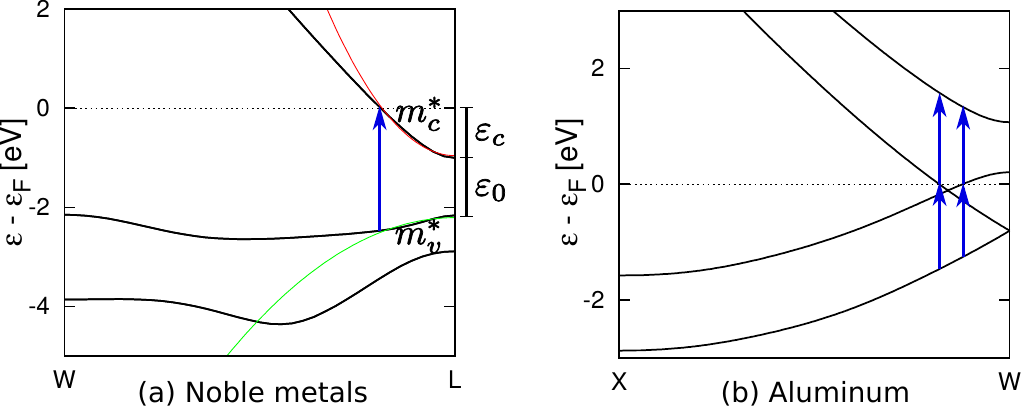}
\caption{
Critical interband transitions determining the `sharp' features
in the dielectric function change for (a) noble metals (gold shown;
similar shapes for silver and copper) and (b) aluminum.
A parabolic band model around the L point
(parameters in Table~\ref{tab:EpsParameters})
approximates the critical transition in noble metals.
This is difficult in aluminum because of four such transitions
in a narrow energy range $\approx 1.3 - 1.6$~eV.
\label{fig:transitionModel}}
\end{figure}

The strongest temperature dependence in noble metals results from
transitions between the highest occupied $d$-band to the Fermi level
near the $L$ point, as shown in Figure~\ref{fig:transitionModel}(a).
Assuming a parabolic dispersion and a constant transition matrix element,
this temperature dependence can be modeled as\cite{ParabolicBandModel,Sun}
\begin{equation}
\Delta\epsilon(\omega) = -\Delta \mathcal{K} \frac{A_0}{(\hbar\omega)^2}\int_{-\varepsilon_c}^\infty
	\frac{d\varepsilon(1-f(\varepsilon,T_e))}{\sqrt{
	{\scriptsize \begin{array}{r}
		(m^\ast_v/m^\ast_c)(\hbar\omega-(\varepsilon+\varepsilon_0+\varepsilon_c))\\
		- (\varepsilon+\varepsilon_c)
	\end{array}}
}}.
\label{eqn:EpsChangeModel}
\end{equation}
The denominator captures the joint density of states
for transitions between the bands, and the numerator counts unoccupied
states near the Fermi level, which introduces the temperature dependence.
Above, $\mathcal{K}$ fills in the real part of the dielectric function,
given the imaginary part using the Kramers-Kronig relation.

\begin{table}
\caption{
Parameters to describe the change in dielectric function of noble metals with
electron temperature using the $d\rightarrow s$ model (\ref{eqn:EpsChangeModel})
with a parabolic band approximation, extracted from fits to the PBEsol+$U$ bandstructure.
The energies and effective masses are also labeled in Figure~\ref{fig:transitionModel}(a).
\label{tab:EpsParameters}}
\setlength{\tabcolsep}{9pt}
\begin{tabular}{lccc}
\hline\hline
 & Ag & Au & Cu \\
\hline
\multicolumn{4}{l}{Physical constants:}\\
$\omega_p$~[eV/$\hbar$] & 8.98 & 9.01 & 10.8\\
$\tau^{-1}$~[eV/$\hbar$] & 0.0175 & 0.0240 & 0.0268\\
\hline
\multicolumn{4}{l}{Fits to DFT calculations:}\\
$A_0$ [eV$^{3/2}$]   &   70  &   22  &  90   \\
$\varepsilon_c$~[eV] &  0.31 &  0.96 &  0.98 \\
$\varepsilon_0$~[eV] &  3.36 &  1.25 &  1.05 \\
$m^\ast_v/m^\ast_c$  &   5.4 &   3.4 &  16.1 \\
\hline\hline
\end{tabular}
\end{table}

Table~\ref{tab:EpsParameters} lists the parameters for the parabolic band
approximation obtained from the PBEsol+$U$ band structures.
Figure~\ref{fig:EpsChange400K} shows that this approximation (dashed lines) captures the
correct shape of $\Delta\varepsilon(\omega)$ for small changes in $T_e$.
However, this model underestimates the $T_e$ dependence for higher
$T_e$ because it ignores the quadratic increase in broadening of
the electronic states due to increased electron-electron scattering,
as Figures~\ref{fig:EpsChange1000K} and \ref{fig:EpsChange5000K} show.
Aluminum exhibits a sharp change in the dielectric function around
$\hbar\omega \approx 1.5$~eV, which results from several transitions
to/from the Fermi level near the W point as Figure~\ref{fig:transitionModel}(b) shows.
Additionally two of the involved bands are not parabolic, making it
difficult to construct a simple model like (\ref{eqn:EpsChangeModel}).
Therefore, simplified models are adequate for qualitative analysis
of lower temperature excitation experiments in noble metals,\cite{Sun}
but dielectric functions from first-principles DFT calculations are necessary for a quantitative analysis
of higher temperature experiments and a wider range of materials and probe frequencies.

\section{Conclusions}
\label{sec:Conclusions}

Our parameter-free DFT calculations of electron-phonon coupling, electron and lattice heat capacities,
and dielectric functions show qualitative differences from free-electron and
previous semi-empirical estimates because of the substantial energy dependence
of electron-phonon matrix elements and electronic density of states.
These changes are particularly important for gold and copper at transient electron
temperatures greater than 2000~K because of the change in occupations of the
$d$-bands situated $\sim 2$~eV below the Fermi level in these metals.

The temperature dependence of the optical response is, in particular,
important for a wide range of applications beyond understanding ultrafast measurements.
We show that while simple models can account for some of the qualitative features
of the change in dielectric function for small changes in temperature,
an electronic structure treatment is essential to quantitatively account for
the complete frequency and temperature dependence, including effects
such as carrier linewidth broadening and transitions between multiple
non-parabolic bands.
Given the dearth of published temperature-dependent dielectric functions
in the literature, we include detailed tables of our
predictions for electron temperatures up to 8000~K, and spanning frequencies
from the infrared to the ultraviolet, in the supplementary information.\cite{NoteSI}

This work has direct implications for analysis of experimental pump-probe
studies of metal nanostructures and is the subject on ongoing work in our group.
With the predicted material properties we anticipate a parameter-free description of
the spectra obtained in transient absorption studies since 
we implicitly account for all the microscopic
processes in the non-equilibrium dynamics of electrons in plasmonic metals.

\section*{Acknowledgements}

This material is based upon work performed by the Joint Center for Artificial Photosynthesis,
a DOE Energy Innovation Hub, supported through the Office of Science
of the U.S. Department of Energy under Award Number DE-SC0004993.
This research used resources of the National Energy Research Scientific Computing Center,
a DOE Office of Science User Facility supported by the Office of Science 
of the U.S. Department of Energy under Contract No. DE-AC02-05CH11231.
The authors acknowledge support from NG NEXT
at Northrop Grumman Corporation.
P. N. is supported by a National Science Foundation Graduate Research Fellowship
and by the Resnick Sustainability Institute.
A. B. is supported by a National Science Foundation Graduate Research Fellowship,
a Link Foundation Energy Fellowship, and the DOE `Light-Material Interactions
in Energy Conversion' Energy Frontier Research Center (DE-SC0001293).

\bibliographystyle{apsrev4-1}
\makeatletter{} 

\end{document}